\documentclass[journal]{IEEEtran}
\usepackage{tikz}
\usepackage{amsmath}
\usepackage{subfigure}
\usepackage{amsthm}

\usepackage{tabularx}
\usepackage{pifont}
\usepackage{flushend}
\usepackage{graphics}
\usepackage{filecontents}
\usepackage{amsmath,amssymb,amsfonts}
\usepackage{url}
\usepackage{caption, subcaption}
\usepackage{epstopdf}
\usepackage{color}
\usepackage{graphicx}

\definecolor{seagreen}{rgb}{0.18, 0.55, 0.34}
\definecolor{royalpurple}{rgb}{0.47, 0.32, 0.66}
\definecolor{brown(traditional)}{rgb}{0.59, 0.29, 0.0}
\definecolor{blue}{rgb}{0.3, 0.2, 0.9}
\definecolor{lightblue}{rgb}{0.93,0.95,1.0} 
\definecolor{lightgreen}{rgb}{0.90,1.0,0.90} 
\definecolor{lightgray}{rgb}{0.83,0.83,0.83} 
\usepackage[colorlinks,
            linkcolor=blue,
            anchorcolor=blue,
            citecolor=blue]{hyperref}

\ifCLASSINFOpdf
\else
\fi
\begin{document}
%
\title{Defining Problem from Solutions: Inverse Reinforcement Learning (IRL) and Its Applications for Next-Generation Networking}
%
%
%

\author{Yinqiu~Liu,
      Ruichen~Zhang,
      Hongyang~Du,
      Dusit~Niyato,~\IEEEmembership{Fellow,~IEEE},
      Jiawen~Kang,
      Zehui~Xiong,~and
      Dong~In~Kim,~\IEEEmembership{Fellow,~IEEE}
      

  \thanks{Y. Liu, R. Zhang, H. Du, and D. Niyato are with the School of Computer Science and Engineering, Nanyang Technological University, Singapore (e-mail: yinqiu001@e.ntu.edu.sg, ruichen.zhang@ntu.edu.sg, hongyang001@e.ntu.edu.sg, and dniyato@ntu.edu.sg).}
  \thanks{J. Kang is with the School of Automation, Guangdong University of Technology, China (e-mail: kavinkang@gdut.edu.cn).}
  \thanks{Z. Xiong is with the Pillar of Information Systems Technology and Design, Singapore University of Technology and Design, Singapore (e-mail: zehui xiong@sutd.edu.sg).}
  \thanks{D. I. Kim is with the Department of Electrical and Computer Engineering, Sungkyunkwan University, Suwon 16419, South Korea (e-mail:dikim@skku.ac.kr).}
  }
\maketitle

\begin{abstract}
%
%
%
%
Performance optimization is a critical concern in networking, on which Deep Reinforcement Learning (DRL) has achieved great success. 
Nonetheless, DRL training relies on precisely defined reward functions, which formulate the optimization objective and indicate the positive/negative progress towards the optimal. 
With the ever-increasing environmental complexity and human participation in Next-Generation Networking (NGN), defining appropriate reward functions become challenging. 
In this article, we explore the applications of Inverse Reinforcement Learning (IRL) in NGN. 
Particularly, if DRL aims to find optimal solutions to the problem, IRL finds a problem from the optimal solutions, where the optimal solutions are collected from experts, and the problem is defined by reward inference. 
Specifically, we first formally introduce the IRL technique, including its fundamentals, workflow, and difference from DRL. 
Afterward, we present the motivations of IRL applications in NGN and survey existing studies. 
Furthermore, to demonstrate the process of applying IRL in NGN, we perform a case study about human-centric prompt engineering in Generative AI-enabled networks. 
We demonstrate the effectiveness of using both DRL and IRL techniques and prove the superiority of IRL. 
\end{abstract}


%
\IEEEpeerreviewmaketitle

\section{Introduction}
Deep Reinforcement Learning (DRL) has become indispensable in many fields, such as networking, robotics, and finance \cite{DRLsurvey}. 
Following the Reinforcement Learning (RL) principle, an agent can optimize its decision-making capability by iteratively interacting with an environment, aiming to maximize cumulative rewards. 
Meanwhile, Deep Neural Networks (DNNs) enhance RL by enabling agents to represent complex environments and learn sophisticated policies.
Although such a paradigm demonstrates remarkable success, explicitly defining or even determining rewards can be challenging in many real-world scenarios \cite{10096237} due to environmental complexity, human participation, or information asymmetry. 
Take task offloading in mobile edge networks as an example, where users select edge servers to maximize their Quality of Experience (QoE).
We know that several factors are related to QoE, such as service latency and fees.
However, the weight and relationship of observed factors, as well as how to fuse them appropriately to model user-side experience, are unknown.
Moreover, the unobserved subjectivity of users, such as personal inner and behavioral preferences, should also be considered.
Without knowing what constitutes the reward, a trained DRL policy might exhibit severe sub-optimality and bias.

Fortunately, Inverse Reinforcement Learning (IRL) emerges as a pivotal solution to overcome the obstacles caused by reward inaccessibility \cite{9844128}.
Specifically, IRL enhances DRL by introducing reward inferences.
In the above example, instead of manually defining a reward function without any accurate prior knowledge and precision guarantee, IRL utilizes DNNs to infer the rewards that can effectively explain user behaviors.
Such behaviors are called expert trajectories and should be collected before training IRL.
Successful IRL applications include ChatGPT, which involves large-scale human feedback to fine-tune model generation\footnote{https://cgnarendiran.github.io/blog/chatgpt-future-of-conversational-ai/}.
Meanwhile, IRL has been widely adopted in human-in-the-loop networking and systems, such as autonomous driving assistance \cite{ARORA2021103500}.
We conclude that IRL owns the following benefits.  
\begin{itemize}
    \item \textbf{Environment Exploration}: IRL provides a means to break the information asymmetry and explore complex or adversarial environments. By leveraging the inferred reward functions, agents are not only guided toward optimal policies but are also encouraged to explore uncharted territories within the environment. For instance, users collect malicious servers' behaviors to infer their objectives and patterns, thus taking corresponding defenses \cite{9844128}.
    \item \textbf{Behavior Understanding}: By inferring reward functions from observed expert behaviors, IRL offers profound insights into the underlying motivations of agents, enabling a deeper comprehension of complex behaviors. For example, human driving patterns are complicated since actions taken by drivers according to different traffic conditions are based on their empirical experience, driving skills, and preferences. Leveraging DNNs to represent such non-linear features, IRL can learn a reward function that aligns with the driving behaviors\footnote{https://meixinzhu.github.io/project/irl/}.
    \item \textbf{Policy Imitation}: IRL excels in distilling policies from demonstrations, allowing the agents to imitate desired behaviors. This capability is especially beneficial in scenarios where the desired outcome is known but the path to achieving it is not. In the driving example, after acquiring a reward function, an autonomous driving agent can be trained to imitate the expert drivers' behaviors. 
\end{itemize}

In this article, we delve into the applications of IRL in Next-Generation Networking (NGN), which is the foundation of numerous advanced technologies, e.g., 6G communications, Metaverse, and Internet of Everything. 
According to AT\&T\footnote{https://www.business.att.com/learn/tech-advice/what-is-next-generation-network-ngn-technology-explained.html}, NGN is projected to accommodate massive devices, support diverse communication/network protocols, and provide immersive services to users.
Such explosive growth in networking scale, topology, and complexity greatly increases the difficulty of defining optimization objectives and rewards precisely.
Noticing such issues, several studies \cite{10096237, 10096376} adopted IRL to solve certain networking problems, such as workload balancing, routing schedules, and attack detection, while an in-depth analysis of IRL-enabled NGN is missing. 
We conclude that three key questions are still yet to be answered.
\begin{itemize}
    \item \textbf{Q1}: How can IRL help NGN, and in which ways?
    \item \textbf{Q2}: How to design IRL algorithms to serve specific NGN scenarios?
    \item \textbf{Q3}: What are the open issues and future directions in this topic?
\end{itemize}
To this end, we provide forward-looking research on IRL-enabled NGN, as well as conducting a case study for demonstration.
\textit{To the best of our knowledge, this is the first work answering why IRL is essential for NGN and showing how to deploy IRL algorithms to solve real-world NGN problems.}
Our main contributions are summarized as follows.
\begin{itemize}
    \item We comprehensively discuss the fundamentals of IRL. Specifically, based on Markov Decision Process (MDP) and RL principles, we introduce the basics of IRL and several representative IRL algorithms. Additionally, we analyze the benefits of enhancing DRL by IRL.
    \item We explore the applications of IRL in NGN. To do so, we first illustrate the driving forces existing in NGN to promote IRL adoption. Afterward, we review the existing literature on IRL-enabled NGN, as well as the limitations of the current IRL techniques.
    \item We conduct a case study to demonstrate the process of applying IRL in NGN. In particular, we showcase a human-centric prompt engineering scheme for Generative AI (GAI)-enabled networks. Experimental results prove that IRL can effectively infer unobserved rewards of human users and achieve higher experience. 
\end{itemize}
\begin{figure*}[!t]
\centering
\includegraphics[width=\textwidth]{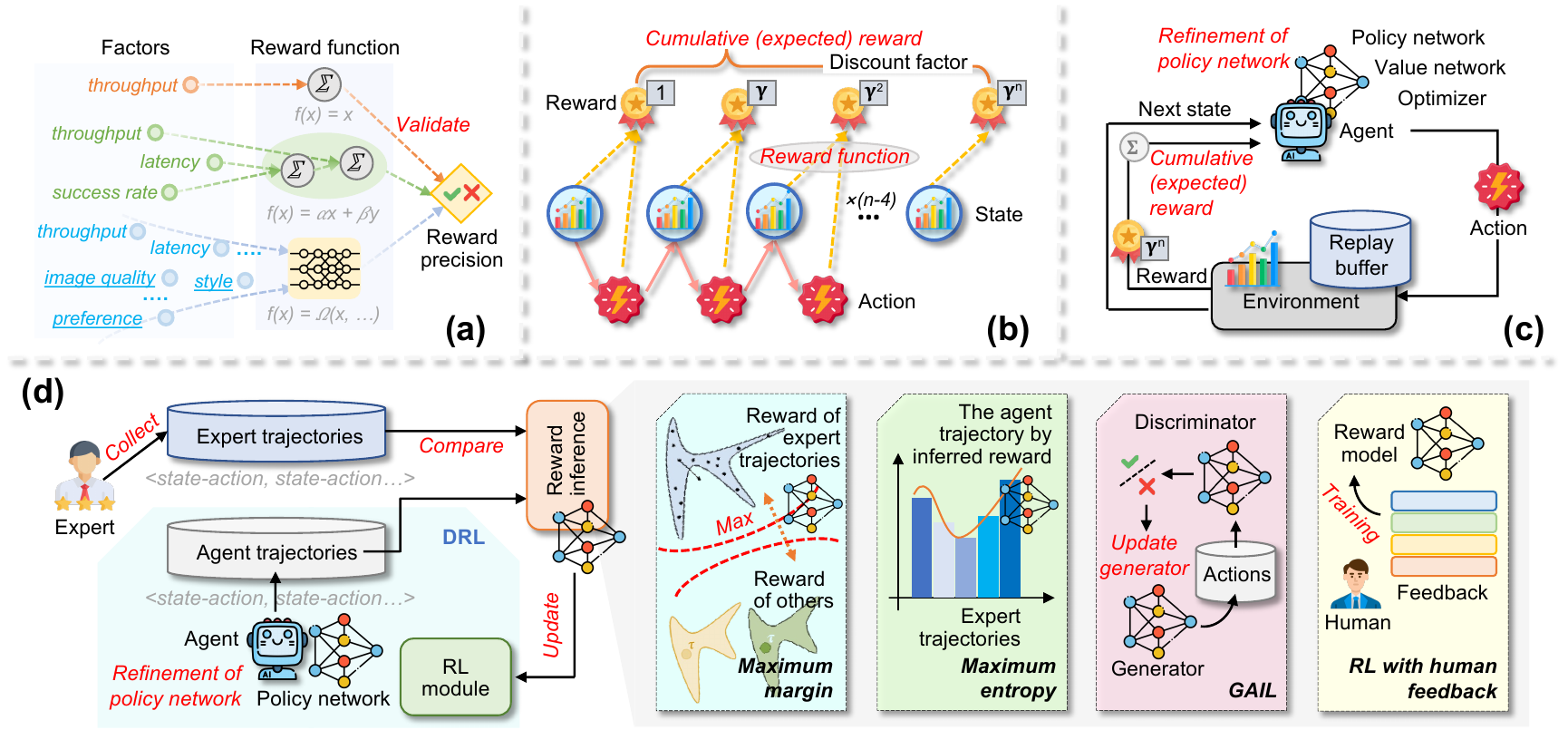}
\caption{The fundamentals of IRL \textbf{(a)}: The cases of defining reward functions. \textbf{(b)}: The illustration of MDP. \textbf{(c)}: The workflow of DRL algorithm. \textbf{(d)} The workflow of IRL. Note that we illustrate four representative approaches to infer reward, namely maximum margin, maximum entropy, GAIL, and RLHF.}
\label{prelim}
\end{figure*}

\section{Fundamentals of Inverse Reinforcement Learning}
In this section, we comprehensively discuss the fundamentals of IRL, including MDP, conventional DRL, the basics of IRL, representative IRL algorithms, and IRL's advantages.

\subsection{Preliminaries}

\subsubsection{Reward Engineering}
To precisely formulate an optimization problem, the reward function should be defined appropriately.
The term ``reward" indicates the desirability of each action.
By assigning positive/higher and negative/lower reward values to the actions that make positive and negative progress, respectively, the optimization objective can be described and the optimal can be gradually approached.
Finally, the inputs of reward functions are observed/unobserved factors of the environments and stakeholders.

As shown in Fig. \ref{prelim}(a), users may encounter different situations when defining reward functions.
First, if their objective is straightforward, such as minimizing the throughput, it can be directly adopted as the reward.
If multiple physical factors, e.g., throughput, latency, and package loss, are involved, the reward function should fuse them in linear/non-linear manners and acquire a combined expression.
Note that the difficulty and potential errors grow dramatically with the increasing number of factors.
Furthermore, in human-in-the-loop networks, the unobserved subjective factors of users should be modeled and considered as well, which is challenging.
The precision of reward functions is measured by how well they can describe optimization objectives and elicit desired behaviors.

\subsubsection{Markov Decision Process}
Fig. \ref{prelim}(b) illustrates an MDP \cite{DRLsurvey}, the building block for DRL and IRL.
As a discrete-time stochastic control process, MDP presents a mathematical framework for modeling sequential decision-making in environments with uncertainty. 
The basic MDP is constructed by a four-element tuple $\langle$\textit{states, actions, transition probabilities, rewards}$\rangle$. 
These components describe the environment in which an agent performs certain actions, acquires rewards, and transits from the current state to the next state.
\begin{itemize}
    \item \textbf{States}: The set of all possible situations, each of which encapsulates various configurations or factors that describe the current environment.
    \item \textbf{Actions}: For each state, there are actions available to the agent that can change the state.
    \item \textbf{Transition Probabilities}: The likelihood of moving from one state to another, given an action taken by the agent.
    \item \textbf{Rewards}: After taking an action and moving to a new state, the agent will receive a reward, whose meaning is discussed before. The reward function can be defined by mathematical expressions or be inferred by DNNs, which correspond to DRL and IRL, respectively. 
\end{itemize}
In the NGN context, most of the optimization problems can be modeled as an MDP, such as resource allocation, routing, attacks/defenses, and sub-carrier selection \cite{DRLsurvey}. 
Hence, acquiring a reliable and effective method for MDP optimization is of significant importance in promoting NGN.

\subsection{Deep Reinforcement Learning}
Combining reinforcement learning (RL) principles with the representational capabilities of DNNs, DRL is the most famous and efficient approach to solving complex decision-making tasks formulated as MDP. 
Next, we introduce the basic idea, architecture, and process of DRL.

\subsubsection{Basic Idea}
The fundamental goal of DRL is to derive an optimal policy that maximizes the cumulative reward in an MDP setup \cite{DRLsurvey}. 
Through iterative interaction with the environment, an agent learns to refine its decision-making strategy based on received feedback. 
DNNs play a crucial role in this process, approximating the functions that predict the future rewards of actions, thereby guiding the agent toward optimal decision-making.

\subsubsection{General Architecture}
As shown in Fig. \ref{prelim}(c), the general DRL framework consists of several essential components:
\begin{itemize}
    \item \textbf{Agent:} The decision-maker that interacts with the environment, i.e., taking actions, transiting to the next state, and receiving rewards.
    \item \textbf{Environment:} The system that defines the state and action spaces and allows agents to perform MDP.
    \item \textbf{Policy Network:} A neural network that maps states to actions, defining the agent's behavior.
    \item \textbf{Value Network:} A neural network that estimates future rewards from states or state-action pairs.
    \item \textbf{Replay Buffer:} A repository of numerous past trajectories, enabling the agent to reuse historical training data and reduce training costs.
    \item \textbf{Optimizer:} The mechanism that adjusts neural network parameters to maximize expected cumulative rewards under the current policy.
\end{itemize}
\begin{figure*}[tpb]
\centering
\includegraphics[width=\textwidth]{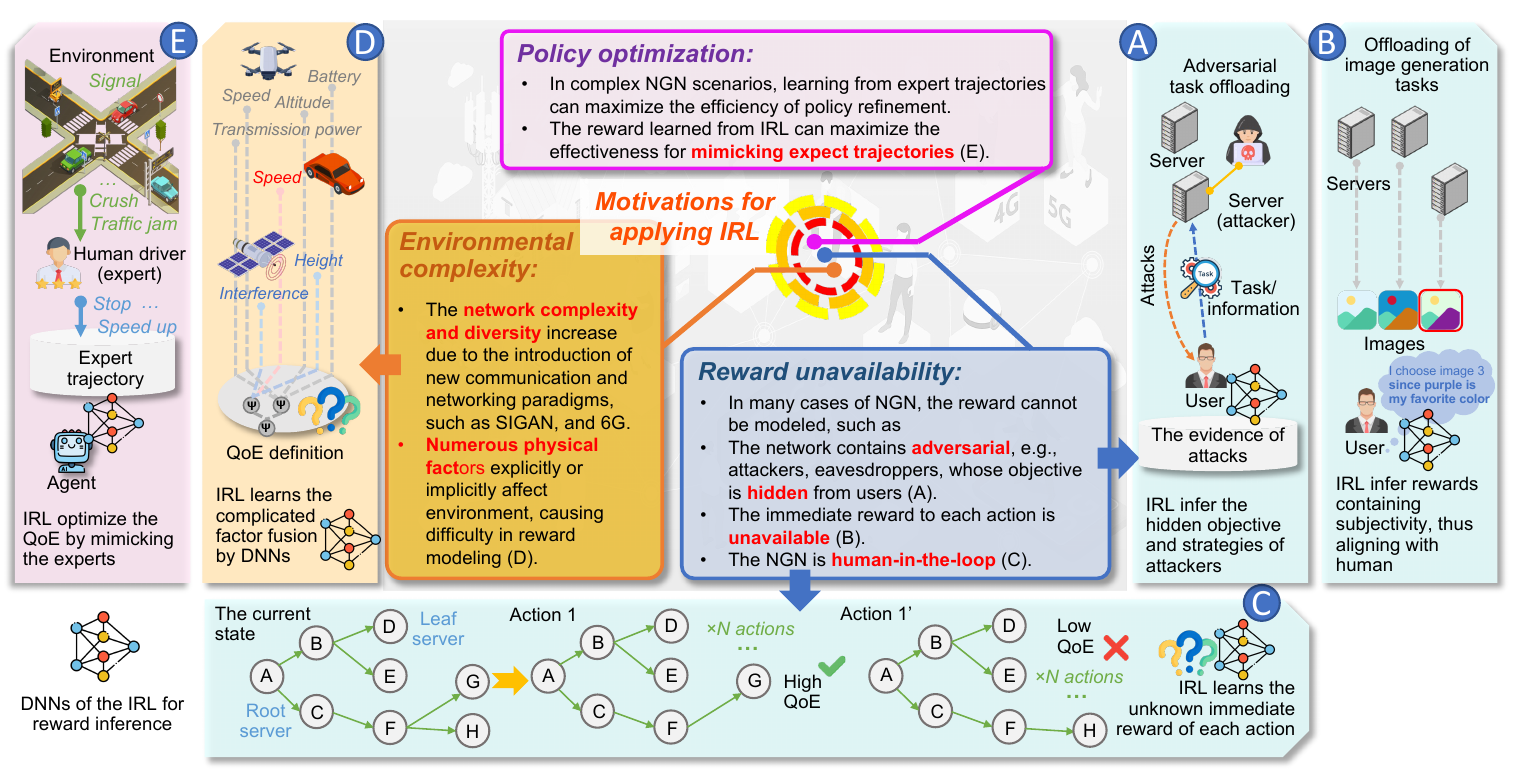}
\caption{The motivation of applying IRL in NGN, using QoE optimization for task offloading as an example. \textbf{A}: The attackers' information is hidden from users. \textbf{B}: Subjectivity factors can hardly be represented precisely. \textbf{C}: The immediate reward of each action is unobserved. Action 1 and 1' choose servers $H$ and $G$ for task offloading, respectively. \textbf{D}: SAGIN communications involve numerous physical factors from diverse devices, causing huge difficulty in QoE modeling. \textbf{E}: The human driving behaviors contain complicated physiological processes that can hardly be described by manually designed rewards.}
\label{NGN}
\vspace{-0.2cm}
\end{figure*}

\subsubsection{Algorithm Process}
To illustrate the DRL process, we show an example of leveraging Proximal Policy Optimization (PPO) \cite{10032267} to offload tasks.
In this case, \textit{agent}, \textit{action}, and \textit{state} refer to the user, the selection of task offloading server, and the edge network, respectively.
In addition, \textit{reward} can be defined by fusing multiple factors, such as latency and success rate, thereby indicating user QoE.
Afterward, the PPO process is streamlined into three main stages, namely data collection, advantage estimation, and policy optimization. 
Initially, the agent interacts with the edge network using the current server selection policy and gathers $\langle$\textit{action, state, reward}$\rangle$ records. 
Such records are buffered to calculate advantage estimates, which assess the relative value of actions. 
Then, PPO optimizes a specially designed objective function that includes a clipping mechanism. 
This mechanism prevents the new policy from straying too far from the old one, ensuring small, stable updates and avoiding drastic performance fluctuations. 
By iteratively cycling through these stages, PPO gradually fine-tunes the policy, striking a balance between exploring new strategies and exploiting known rewards, making it adept at navigating complex environments.

\subsection{Inverse Reinforcement Learning}
Despite the success of DRL, in many network scenarios, precisely defining the reward function is impossible.
In the above example, although manually-defined QoE can reflect users' preference for low latency and high success rate, the subjectivity factors are ignored.
For instance, users with time-sensitive tasks may emphasize latency, while secure computing tasks should put success rate at the highest priority. 
Following biased reward functions, the policy training may deviate from the real optimization objective.
Accordingly, IRL is presented to infer such sophisticated reward functions by observing and analyzing users' behaviors \cite{10096237}.
As shown in Fig. \ref{prelim}(d), in the beginning, the uses' selections of edge servers in different network states are collected, called expert datasets/trajectories. 
Afterward, instead of simply learning the action policy, IRL first infers the form of the reward function that best explains the demonstration behaviors.
Moreover, it can optimize the agent's policy towards closely aligning with or mimicking the expert policy according to the inferred reward function.
Given the strong representation and learning ability, the reward function is generally inferred by a DNN. 

\subsection{Evolution of Inverse Reinforcement Learning}
The inference of reward functions can be implemented following different approaches.
As illustrated in Fig. \ref{prelim}(c), we introduce four representative and widely adopted IRL algorithms, from basic maximum margin to advanced generative imitation learning.
\subsubsection{Maximum Margin} This algorithm \cite{ARORA2021103500} focuses on distinguishing the expert's policy from all other policies by a margin that depends on the difference in the accumulated reward. It operates under the principle that the correct reward function should make the expert behavior appear significantly better than all other possible behaviors. This method is particularly useful in scenarios where clear demarcation between optimal and suboptimal policies is possible.
\subsubsection{Maximum Entropy} This method \cite{ARORA2021103500} introduces the concept of \textit{entropy} to address the ambiguity in the reward function that could explain observed behaviors. It assumes that among all possible reward functions, the one that leads to the observed behavior while also maximizing entropy, or in other words, promoting diverse but consistent behaviors, is the most appropriate. Compared with maximum margin, this approach is adept at handling the inherent uncertainty in determining why an agent acts a certain way by favoring reward functions that support a wide range of plausible actions.
\subsubsection{Generative Imitation Learning} Recall that reward functions serve as an indicator to guide policy refinement. Apart from inferring an explicit reward value for each action and maximizing the reward expectation, the policy can be refined in an imitation manner. For instance, GAIL \cite{ARORA2021103500} leverages the Generative Adversarial Networks (GANs) framework, where a DNN-based generator aims to approach expert behavior. Meanwhile, another DNN acts as the discriminator, trying to differentiate between the expert's actions and those of the generator. The two modules are trained alternately, i.e., freezing one module and adjusting the parameters of the other module to optimize its behavior imitation/differentiation performance. Afterward, the generator can effectively mimic the expert policy and solve MDP problems without requiring reward inference. Accordingly, the computational efficiency can be significantly improved. 
\subsubsection{Reinforcement Learning with Human Feedback} In many cases, the reward function is unavailable since reward values are given by humans. Human perception and evaluation is a complex physiological process and is affected by various subjective factors such as preference, personality, environment, and strictness. As a variant of IRL, RLHF combines the RL principle with direct feedback from humans, integrating subjective insights into the learning process. Specifically, a reward model will be first trained based on action-score pairs, where scores are manually annotated by humans. Then, the DRL can be leveraged for policy optimization, ensuring the policy aligns with human preferences.

\subsection{Comparison and Summary}
Based on the above descriptions, the difference between DRL and IRL can be summarized as follows.
First, DRL operates within a well-defined MDP environment with available four-element tuples. 
IRL, conversely, applies to incomplete MDPs, where the reward function is unknown.
Accordingly, the primary goal in DRL is to learn an optimal policy that maximizes cumulative rewards. 
IRL aims to infer the reward function based on the expert's demonstration to understand the objectives the agent is implicitly striving for.
The inferred reward then guides the policy refinement.
Finally, the DRL policies are optimized by iteratively interacting with the environment. 
In contrast, IRL relies on offline expert demonstrations, learning indirectly from optimal behaviors and inferring the rewards that can elicit such behaviors.
\textit{Overall, DRL is used to find a solution to the problem while IRL is used to find a problem from the solutions}.

The most significant advantage of IRL is its ability to infer unobserved reward functions precisely. 
Such capability greatly enhances conventional DRL since if the reward is defined inappropriately, the trained policy cannot solve the optimization efficiently.
In addition, IRL, especially RLHF, excels in understanding and mimicking human-like decision-making behaviors, making it ideal for applications requiring nuanced behavior modeling. 
Last but not least, IRL can provide insights into the underlying motivations and objectives behind observed actions, contributing to understanding environments.
\begin{table*}[tpb]
\centering
\caption{Summary of applications of IRL in networking.} 
\includegraphics[width=\textwidth]{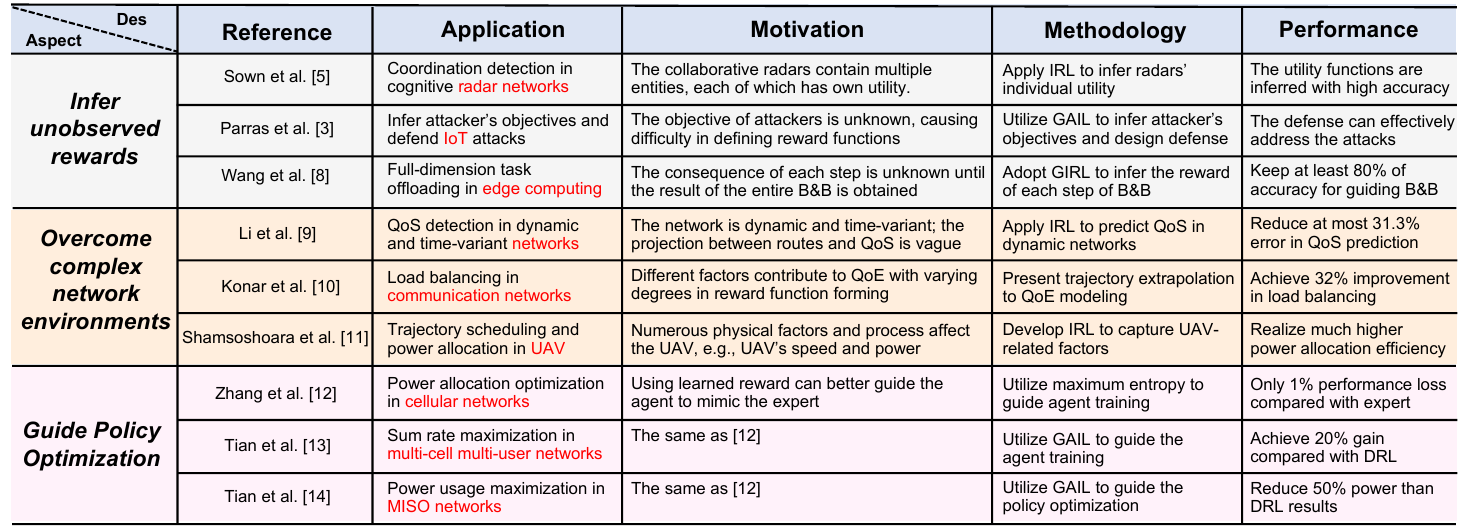}
\label{survey}
\end{table*}

\section{Inverse Reinforcement Learning in Next-Generation Networking}
\subsection{Motivations from Next-Generation Networking}
According to AT\&T\footnote{https://www.business.att.com/learn/tech-advice/what-is-next-generation-network-ngn-technology-explained.html}, NGN refers to the evolution and migration of fixed and mobile networking infrastructures from distinct and proprietary networks to converged networks with high efficiency, security, and experience. 
Hence, the manners of networking organization, management, and service provisioning will undergo significant changes \cite{NGN2}.
In this part, we utilize QoE maximization for task offloading as an example to demonstrate the driven force in NGN that promotes the further development of IRL (see Fig. \ref{NGN}).
\begin{itemize}
    \item \textbf{Reward Unavailability}: In some NGN scenarios, rewards are unavailable to decision-makers. As shown in Fig. \ref{NGN}(a), users may encounter malicious offloading servers, which steal sensitive information for attacks. The intelligent defense against such malicious behaviors relies on detecting attackers' objectives, which are hidden from users. In addition, the complicated topology of NGN leads to sophisticated RL state spaces \cite{NGN2}. For instance, Fig. \ref{NGN}(b) illustrates a case where offloading servers are structured as a tree and users choose an offloading service chain from the leaf to the root. In this case, the immediate reward of performing each action is inaccessible since the association between the current state and the final result is unknown. Finally, NGN exhibits human-in-the-loop features since it aims to support diverse advanced applications and ensures high human-perceptual experiences \cite{NGN2}. Take the offloading of image generation tasks as an example (see Fig. \ref{NGN}(c)), user QoE depends not only on objective factors such as service latency and fee but also on users' preference for the painting style, whose modeling is intricate.
    \item \textbf{Environmental Complexity}: Given massive devices, as well as the advancement of access protocols and communication diagrams, NGN exhibits environmental complexity. For instance, in Space-Air-Ground Integrated Network (SAGIN), multiple physical factors of the devices in each layer explicitly or implicitly contribute to QoE, while the specific contribution is unknown (see Fig. \ref{NGN}(d)). In this case, the annually defined reward function may suffer from incomprehensiveness and bias.
    \item \textbf{Policy Optimization}: If expert behaviors that maximize QoE are available, training the agent to mimic the expert can lead to the highest efficiency for QoE optimization. However, manually designing a reward function can hardly realize the action imitation. Fig. \ref{NGN}(e) demonstrates an example of autonomous driving. Contributed to the strong capability of DNNs, IRL can mimic the expert driver with perfect QoE effectively by inferring the reward that elicits desired behaviors.
\end{itemize}

\subsection{Applications of Inverse Reinforcement Learning in Next-Generation Networking}
In this part, we review related works about IRL in networking, as shown in TABLE I.
Particularly, our survey is organized from the three perspectives shown above.
\subsubsection{Infer Unobserved Rewards} As mentioned in Section II, IRL is effective in inferring unobserved rewards caused by adversarial relationships or human participation in the networks. 
For instance, Snow \textit{et al.} \cite{10096376} applied multi-agent IRL to identify coordination in cognitive radar networks, i.e., multiple radars collaborate to track one target. 
Since the radars are adversarial, the target cannot identify their numbers, configurations, and individual utility functions. 
Leveraging the IRL concept, the target first determines if collaboration exists based on whether the emissions satisfy Pareto optimally \cite{10096376}. 
If so, it then uses the emissions as the expert trajectories to infer the individual utility of each radar. 
Likewise, Parras \textit{et al.} \cite{9844128} leveraged GAIL to enhance IoT security. 
This is because, nowadays, DRL is widely adopted by attackers to create new attacks. 
Specifically, they select the attack objective, adopt DRL to train the optimal attack strategy and launch the attacks on the victims. 
Accordingly, defenders can adopt IRL methods, such as GAIL, to infer the objective of the attackers and thus refine the defense policies. 

In addition to adversarial networks, Wang \textit{et al.} \cite{10285035} discussed the applications of IRL in cases where the reward of each action cannot be mathematically calculated. 
Specifically, they utilize the branch and bound (B\&B) algorithm to realize full-dimensional task offloading in edge computing.
Although the B\&B process can be modeled as an MDP, it is very challenging to design an appropriate reward function since the B\&B algorithm.
Before obtaining an optimal solution, the branching order of variables cannot be known in advance. 
That is because the B\&B algorithm solves the problem by breaking it down into smaller sub-problems and using a bounding function to eliminate sub-problems with limited upper bounds, forming a tree-structured solution space.
In this case, we cannot know that branching which variable can generate a smaller enumeration tree.
Therefore, the authors present a Graph-based IRL (GIRL), which uses a Graph Neural Network (GNN) to infer the immediate reward of each action based on the structure of the tree.

Finally, the concept of IRL has been widely adopted in human-in-the-loop networks.
For instance, the training of ChatGPT, the most famous multimodal AIGC model, involves self-reinforcement from large-scale human feedback, thereby aligning the model output with human preferences.

\subsubsection{Overcome Complex Network Environments} 
Owing to complex network environments, even if manually defining some vague reward functions is available, it can hardly precisely represent the environmental change due to the agent's actions and efficiently indicate the desirability. 
In \cite{10356665}, the authors presented an IRL-based approach to predict QoS in dynamic and time-variant networks. 
Due to large state spaces and complex projections between routes and QoS values, it is difficult to define a precise reward function artificially. 
Likewise, Konar \textit{et al.} \cite{10279136} presented the trajectory extrapolation algorithm to model the quality of experience (QoE) for communication load balancing.
Particularly, they consider the difficulty of gathering exert trajectories in practice.
Hence, their proposed algorithm first sorts the gathered trajectories according to some objective and well-established metrics.
Then, the trajectories are sampled to facilitate the reward inference and policy learning.
In UAV networks, Shamsoshoara \textit{et al.} \cite{shamsoshoara2023joint} developed the interference-aware joint path planning and power allocation scheme to minimize the interference that UAVs cause to the existing ground user equipment.
IRL is applied to capture all physical factors, including the UAV’s transmission power, terrestrial user density, imposed interference, and the UAV's task completion capability, thereby inferring the optimization objective. 

\subsubsection{Guide Policy Optimizations} 
In some cases where the optimal trajectories are available, directly inferring the reward function that facilitates the agent to mimic the expert can lead to the best policy optimization efficiency. 
For instance, Zhang \textit{et al.} \cite{9798257} adopted maximum entropy to optimize the power allocation schemes in multi-user cellular networks. 
Experiments show that IRL greatly outperforms manually defined reward functions based on objective metrics, such as weighted minimum mean square error. 
Similarly, Tian \textit{et al.} \cite{10172220} applied this principle to multi-cell multi-user networks to maximize the sum rate. 
Furthermore, in \cite{ICCCS}, they adopted the GAIL method to minimize power usage in multiple-input-single-output networks with multiple users. 
Compared with conventional DRL, the proposed method can reduce power consumption by over 50\%.

\textbf{Lesson Learned}: From the above brief survey, we learn that IRL enhances the capabilities of DRL by introducing a reward inference step. 
Contributed to DNN's stronger representation and learning capabilities than humans, IRL can well overcome complex environments, actions, and reward perception processes (such as human perception of AI-generated images). 
Such advantages make IRL inherently suitable for scenarios that are complex while existing expert trajectories for demonstration.
Nonetheless, IRL also exhibits certain drawbacks, such as the additional computational costs for reward inferences and the dependency on expert trajectories, which are subject to further research.

\begin{figure}[tpb]
\centering
\includegraphics[width=0.5\textwidth]{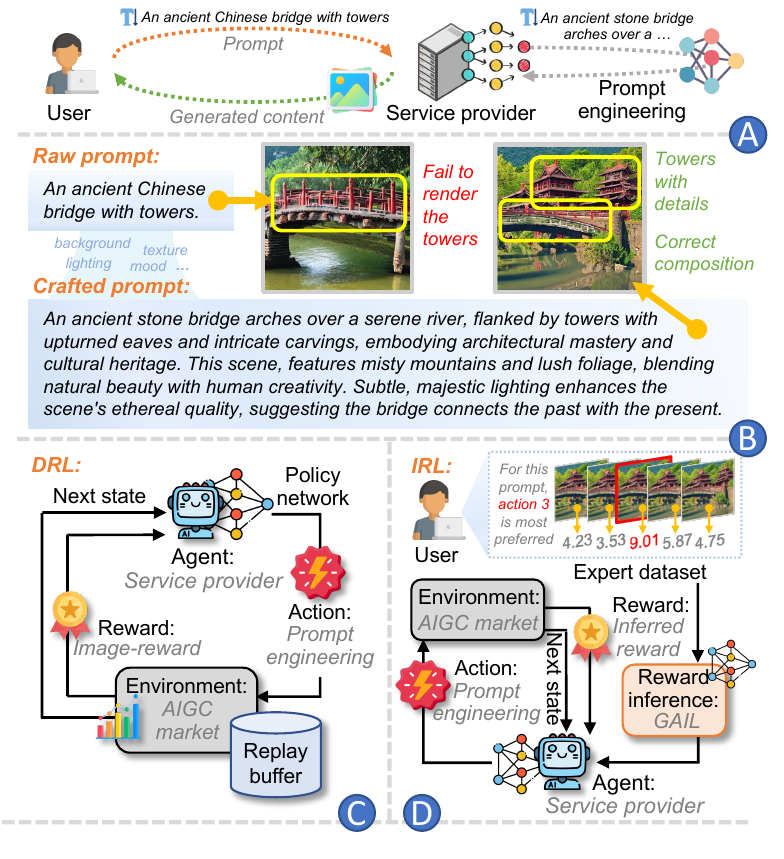}
\caption{The illustration of our case study. \textbf{A}: The system model of GAI-enabled network. \textbf{B}: The efficacy of prompt engineering. We can observe that the image generated from the crafted prompt excels in object rendering and image composition. \textbf{C}: The illustration of DRL-based prompt engineering. \textbf{D}: The illustration of IRL-based prompt engineering.}
\label{GAI}
\end{figure}
\section{Case Study: Human-oriented Prompt Engineering in Generative AI-Enabled Networks}
To illustrate how to solve practical NGN problems by IRL, this section performs a case study about human-oriented prompt engineering.
Particularly, we simultaneously present the DRL- and IRL-based solutions to the same task, helping readers compare these two methods.
The effectiveness of DRL and IRL is also discussed.

\subsection{System Model}
With the rapid advancement of GAI, generative tasks, such as text-to-image generation, automatic conversation, and 3D avatar renderings, play an important role in NGN.
As shown in Fig. \ref{GAI}(A), we consider a GAI-enabled network with users and service providers.
Users describe their requests by natural language (so-called prompt) and submit them to the service providers.
Operating professional GAI models, the service providers perform generative inferences and generate required content for users.

Nonetheless, due to the lack of professionality/experience, user prompts may suffer from information insufficiency and ambiguity, causing low generation quality.
To this end, service providers can strategically craft prompts that maintain semantic equivalence with raw prompts while being informative, clear, and suitable for the specific GAI model \cite{Prompt1}. 
Such a process is called \textit{prompt engineering} \cite{Prompt1}.
Taking text-to-image generation as an example, Fig. \ref{GAI}(B) depicts the impact of prompt engineering on the generated image.
We can observe that the image generated by the crafted prompt outperforms in exquisiteness and composition.
Despite such advantages, service providers may own multiple prompt engineering approaches since the prompts are crafted by open vocabularies.
In this case, how to select the optimal prompt engineering approach according to the specific request is a remaining challenge.
Next, we solve such an optimization problem following DRL and IRL paradigms, respectively.

\subsection{DRL- and IRL-based Prompt Engineering}
\subsubsection{DRL Workflow} First, we solve this problem following the DRL paradigm. 
Specifically, we adopt the PPO algorithm discussed in Section II-B.
As shown in Fig. \ref{GAI}(C), the service provider acts as \textit{agent}, whose \textit{actions} include all available operations to refine raw prompts.
The raw prompts from users are accommodated by \textit{environment}.
Finally, \textit{reward} evaluates the efficacy of applying the selected prompt engineering operation on the current raw prompt.
To this end, we leverage Image-reward\footnote{https://github.com/THUDM/ImageReward}, a learning-based metric for measuring the aesthetic quality of AI-generated images, as the reward function.
Correspondingly, the quality score of the image generated from the crafted prompt is utilized as the reward.
The DRL architecture and algorithm workflow follow the description in Section II-B, with the goal of optimizing the policy for selecting prompt engineering operations.

\subsubsection{IRL Workflow}
Then, we apply IRL to solve the same problem.
Different from DRL, as shown in Fig. \ref{GAI}(d), IRL include the following three steps.
\begin{itemize}
    \item \textbf{Expert Dataset Collection}: First, we construct an expert dataset to indicate human preference for AI-generated images. Particularly, inspired by the outstanding capability of Large Language Models (LLMs) in multimodal understanding \cite{Prompt1}, we leverage an LLM-empowered agent to mimic real users. Then, we randomly compose 50 raw prompts and perform all the available prompt engineering operations on each of them. The agent is utilized to evaluate the generated images by scores. Note that we adopt two strategies for training LLM, ensuring it generates rationale scores. First, we instruct LLM to act as experienced users, recalling its pre-trained knowledge of image quality assessment. Additionally, we feed ten materials regarding computer vision, image generation, and painting to the LLM, thus enhancing its expertise. All the \textit{$\langle$raw prompt, crafted prompt, score$\rangle$} pairs construct the expert dataset, showing which kind of prompt engineering is preferred by humans for the specific image generation task.
    \item \textbf{Policy Optimization}: Different from defining rewards via an existing metric, IRL utilizes DNNs to infer reward functions that align with expert decisions. To improve the learning efficiency, we adopt GAIL explained in Section II-D. Specifically, the discriminator distinguishes between the selections of expert prompt engineering policy and those of the agent's policy, thereby self-calibrating rewards to align with the expert decision-making mode. Meanwhile, the generator aims to learn a prompt engineering policy that mimics the expert behaviors, driven by feedback from the discriminator. In this way, the generator can gradually imitate humans in judging AI-generated images and perform prompt engineering accordingly.
    \item \textbf{MDP Design}: Finally, we configure the MDP for IRL. Similar to DRL, the \textit{action} space contains all candidate prompt engineering operations. The state space is represented by the LLM-assigned scores of the image generated by the crafted prompt. This allows GAIL to evaluate the efficacy of each action by evaluating its impact on the resulting image quality. Unlike traditional DRL which relies on manually designing reward functions, GAIL leverages the discriminator network to infer rewards from our expert dataset. 
\end{itemize}

\subsection{Experiments}
\subsubsection{Experimental Setting}
We take text-to-image generation as an example.
The service providers adopt Stable Diffusion v2 to generate images.
Raw prompts from users take the form of \textit{A [A] with [B]}, e.g., ``a city with a car" and ``a garden with a fountain."
Leveraging the GPT-4 model, seven kinds of operations to craft prompts can be performed on each raw prompt, as discussed in \cite{Prompt1}.
The LLM-based agent is also implemented by GPT-4.
For IRL, the discriminator consists of a four-layer fully connected DNN and two intermediate layers, each containing 100 neurons. 
Meanwhile, the generator containing actors and critics employs a similar four-layer, fully connected DNN architecture with 64 neurons in each of the two hidden layers. 
The hyperbolic tangent function serves as the activation mechanism for all hidden layers. 
The learning rate and gamma are set to $3\times 10^{-4}$ and 0, respectively.

\subsubsection{Result Analysis}

\begin{figure}[tpb]
\centering
\includegraphics[width=0.47\textwidth]{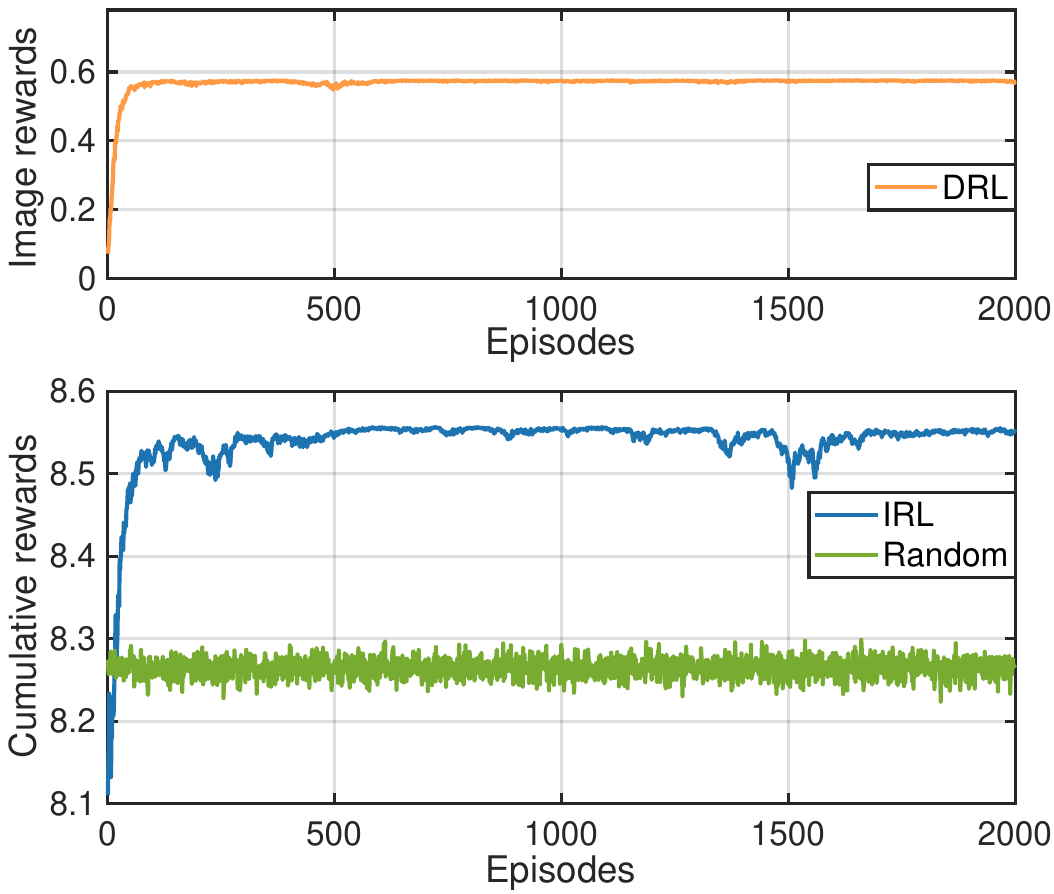}
\caption{The training curves of DRL and IRL.}
\label{IRL}
\end{figure}

\begin{figure}[htpb]
\centering
\includegraphics[width=0.47\textwidth]{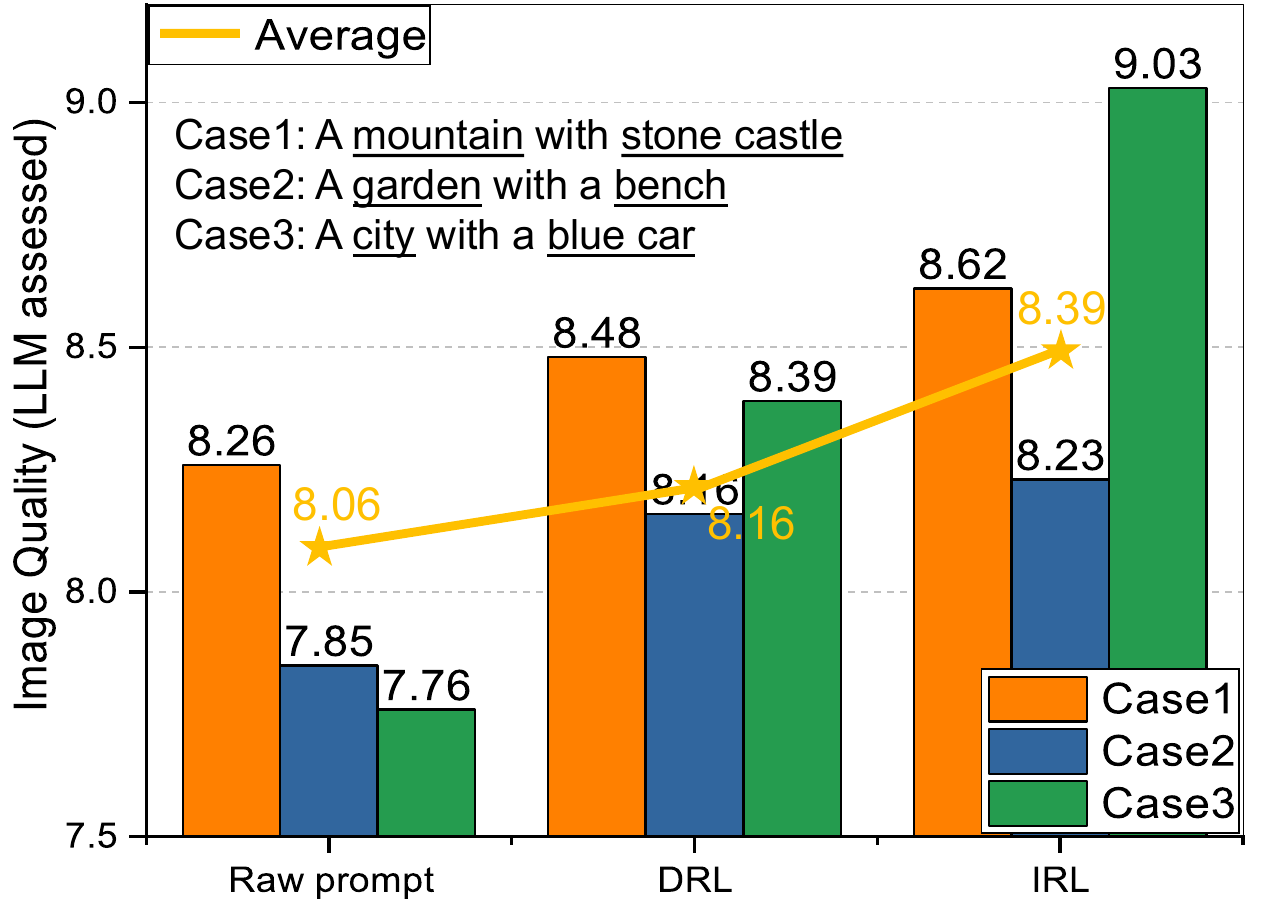}
\caption{The efficacy of DRL and IRL in prompt engineering. Each case corresponds to one randomly selected raw prompt.}
\label{IRL2}
\end{figure}

Fig.~\ref{IRL} shows the trend of cumulative rewards of different algorithms during the training process. 
It shows that as the number of episodes increases, the proposed IRL method achieves good convergence and significantly outperforms the baseline, i.e., the service provider selects prompt engineering operation randomly. 
Meanwhile, DRL also converges rapidly.
With well-trained DRL and IRL policies, Fig. \ref{IRL2} evaluates their efficacy in selecting prompt engineering operations. 
Note that the quality scores are also assessed by the LLM-empowered agent.
As shown in Fig. \ref{IRL2}, IRL can increase the image quality by 0.33 on average, while DRL can only achieve 0.1 increment.
This is because IRL is trained on expert datasets, which can effectively indicate the human preference for assessing AI-generated images.
Consequently, the prompt engineering operations selected by IRL can better align with human desire, resulting in high-score images.

\textbf{Discussion}: Contributed to the increasing image quality, the service experience of humans can be increased drastically.
Meanwhile, the re-generation caused by unqualified outputs can be reduced, which greatly decreases service latency and bandwidth consumption \cite{Prompt1}.


\section{Future Directions}

\subsection{Mixture-of-Experts}
Recall that IRL heavily relies on expert trajectories, while in some complex scenarios, collecting the optimal trajectories that achieve the objective with the lowest cost is impossible.
Instead, there may only exist multiple local optimal trajectories owned by distributed experts, each of which reaches the optimum in certain aspects/stages.
To this end, the mixture-of-experts principle can be leveraged, which utilizes a learning-based gating network to dynamically select/combine different trajectories in different training stages. 

\subsection{IRL with Human Feedback}
Human feedback has been successfully integrated into DRL (e.g., ChatGPT) to make policies align with human preferences.
Likewise, humans can participate in the IRL process to further enhance the expert dataset/human annotation.
For example, in our case study, the efficient calibration of inferred rewards based on human feedback is worth researching, thereby ensuring the reward function represents the real human judgment of AI-generated images precisely.

\subsection{Security of IRL}
The reliance on expert trajectories may cause security issues since attackers can easily mislead policy training by data poisoning.
To this end, strict access control and privacy protection are urgently required for deploying IRL in practical NGN scenarios.
Zero-trust can be a potential technique to dynamically manage data access and usage, thereby preventing privacy leakage.

\section{Conclusion}
In this article, we have explored the applications of IRL in NGN.
Specifically, we have comprehensively introduced the IRL technique, including its fundamentals, representative algorithms, and advantages.
Then, we have discussed the vision of NGN, as well as the motivations for adopting IRL.
Afterward, we have surveyed existing literature about IRL proposals for solving networking problems. 
Furthermore, we have performed a case study on human-centric prompt engineering in GAI-enabled networks, comparing the workflow and effectiveness of both DRL and IRL.
Finally, the future directions to promote the further development of IRL in NGN have been summarized.

\bibliographystyle{IEEEtran}
\bibliography{bare_jrnl}
\vfill

\end{document}